\documentclass[12pt]{article}
\usepackage{amsfonts}

\def\nqq{\hspace{-2em}}
\def\beq#1{\begin{equation}\label{#1}}
\def\eeq{\end{equation}}
\def\ber#1{\begin{eqnarray}\label{#1} \nqq}
\def\eer{\end{eqnarray}}
\def\nn{\nonumber}
\newcommand{\bear}[1]{\begin{eqnarray}\label{#1}}
\newcommand{\ear}{\end{eqnarray}}
\newcommand{\R}{{\mathbb R}}
\newcommand{\N}{{\mathbb N}}

\newcommand{\sign}{\mathop{\rm sign}\nolimits}

\newcommand{\eps}{\varepsilon}
\newcommand{\tri}{\triangle}
\newcommand{\p}{\partial}

\newcommand{\fnm}{\footnotemark}
\newcommand{\fnt}{\footnotetext}

\begin{document}

\thispagestyle{empty}

\begin{flushright}        IGC-PFUR-02/2002  \end{flushright}

\vspace{15pt}

\begin{center}
\Large\bf
BLACK-BRANE SOLUTION FOR $A_3$ ALGEBRA \\[15pt]
\large
M.A. Grebeniuk\fnm[1]\fnt[1]{mag@gravi.phys.msu.su},
V.D. Ivashchuk\fnm[2]\fnt[2]{ivas@rgs.phys.msu.su} and
V.N. Melnikov\fnm[3]\fnt[3]{melnikov@rgs.phys.msu.su} \\[10pt]

\vspace{10pt}

\normalsize\it
Center for Gravitation and Fundamental Metrology, VNIIMS, \\
3/1 M. Ulyanovoy Str., Moscow 117313, Russia and \\[5pt]
Institute of Gravitation and Cosmology, People's Friendship University \\
of Russia, Mikhlukho-Maklaya Str. 6, Moscow 117198, Russia
\end{center}

\vspace{15pt}

\begin{abstract}
Black $p$-brane solutions for a wide class of intersection rules and
Ricci-flat ``internal'' spaces are considered. They are defined up to
moduli functions $H_s$ obeying non-linear differential equations with
certain boundary conditions imposed. A new solution with intersections
corresponding to the Lie algebra $A_3$ is obtained. The functions $H_1$,
$H_2$ and $H_3$ for this solution are polynomials of degree 3, 4 and 3,
correspondingly. An example of $A_3$-solution  with three $3$-branes in
$12$-dimensional model (suggested by N. Khviengia et al) is presented.
\end{abstract}

\hspace{1cm} PACS:
04.20.Jb; 04.50.+h; 04.70.Bw; 02.20.Sv; 02.30.Hq

\hspace{1cm}

\newpage

\section{Introduction}

In this paper we continue investigation of the black-brane solutions with
arbitrary intersection rules \cite{IMp1,IMp2,IMp3,IMtop}. These solutions
are governed by functions $H_s = H_s(R)$ obeying a set of second order
non-linear differential equations (equivalent to Toda-type ones) with
certain boundary conditions imposed.

More general spherically symmetric solutions
were obtained in \cite{IK} using the sigma-model approach from
\cite{IMC,IMJ,IMtop}. For another (algebraic) approach
see also \cite{AR,AIR,AIV} and references therein.

In order to simplify the investigation of such model
and obtain exact solutions the following
conjecture was suggested in \cite{IMp1}: the functions $H_s$ are
polynomials if the $p$-brane intersection rules correspond to semisimple
Lie algebras. This conjecture was verified for
solutions with $A_m$ and $C_m$ series of
Lie algebras \cite{IMp2,IMp3} and also confirmed by special black-brane
"block-orthogonal" solutions considered earlier in \cite{Br,IMJ2}.
It should be noted that an analogue of this conjecture for extremal black
holes was considered earlier in \cite{LMMP}.

Although the conjecture mentioned above allows one to reduce the set of
differential equations to the system of algebraic one, it still seems
to be very difficult to get the general solution to this set of equations
too. So, the only way of obtaining exact solutions
is to investigate special cases with
different Lie algebras (such as $A_2$, $C_2$ etc.).

The simplest nontrivial example of black-brane solution corresponding to
Lie algebra $A_2 = sl(3)$ was obtained in \cite{IMp1,IMp2} (e.g. dyonic
solutions in $11$-dimensional supergravity). This $A_2$-solution was
governed by two polynomials $H_1$ and $H_2$ of degree $2$. The coefficients
of polynomials and charges were the functions of some parameters $P_1$
and $P_2$ and had a rather simple form.

Here we present a new solution corresponding to the Lie algebra $A_3$
and governed by three polynomials of degree 3, 4 and 3. The coefficients
of polynomials exhibit a non-meromorphic dependence upon parameters $P_1$,
$P_2$ and $P_3$: a  term, which is a square root of a polynomial
of degree 4, appears. (For the special case with $P_1 = P_3$
the solution was already obtained and
investigated in \cite{GIK}). Here we also suggest an example of
$A_3$-solution with three $3$-branes in $12$-dimensional model from
\cite{KKLP}.

\section{$p$-brane black hole solutions}

Consider a model governed by the action
\ber{1.1}
S=\int d^Dx \sqrt{|g|}\biggl\{R[g]-h_{\alpha\beta}g^{MN}\p_M\varphi^\alpha
\p_N\varphi^\beta-\sum_{a\in\tri}\frac{\theta_a}{n_a!}\exp[2\lambda_a
(\varphi)](F^a)^2\biggr\}
\eer
where $g=g_{MN}(x)dx^M\otimes dx^N$ is a metric, $\varphi=(\varphi^\alpha)
\in\R^l$ is a vector of scalar fields, $(h_{\alpha\beta})$ is a constant
symmetric non-degenerate $l\times l$ matrix $(l\in \N)$, $\theta_a=\pm1$,
\beq{1.2a}
F^a = dA^a = \frac{1}{n_a!} F^a_{M_1 \ldots M_{n_a}}
dz^{M_1} \wedge \ldots \wedge dz^{M_{n_a}}
\eeq
is a $n_a$-form ($n_a\ge1$), $\lambda_a$ is a 1-form on $\R^l$:
$\lambda_a(\varphi)=\lambda_{\alpha a}\varphi^\alpha$, $a\in\tri$,
$\alpha=1,\dots,l$. In (\ref{1.1}) we denote $|g| = |\det (g_{MN})|$,
$(F^a)^2_g = F^a_{M_1 \ldots M_{n_a}} F^a_{N_1 \ldots N_{n_a}} g^{M_1 N_1}
\ldots g^{M_{n_a} N_{n_a}}$, $a \in \tri$. Here $\tri$ is some finite set.
It should be noted that in the models with one time all $\theta_a = 1$
if the signature of the metric is $(-1,+1, \ldots, +1)$.

Let us consider (black-brane) solutions to the field equations
corresponding
to the action (\ref{1.1})  from \cite{IMp1,IMp2,IMp3}. These solutions are
defined on the manifold
\beq{1.2}
M = (R_{0}, + \infty) \times (M_1 = S^{d_1}) \times (M_2 = \R)
\times  \ldots \times M_n,
\eeq
and have the following form
\bear{2.30}
g= \Bigl(\prod_{s \in S} H_s^{2 h_s d(I_s)/(D-2)} \Bigr)\biggl\{ f^{-1}
dR \otimes dR + R^2  d \Omega^2_{d_1}  \\ \nn
- \Bigl(\prod_{s \in S} H_s^{-2 h_s} \Bigr) f dt \otimes dt
+ \sum_{i = 3}^{n} \Bigl(\prod_{s\in S} H_s^{-2 h_s \delta_{iI_s}}
\Bigr) g^i  \biggr\}, \\ \label{2.31}
\exp(\varphi^\alpha)= \prod_{s\in S} H_s^{h_s \chi_s
\lambda_{a_s}^\alpha}, \\  \label{2.32a}
F^a= \sum_{s \in S} \delta^a_{a_s} {\cal F}^{s},
\ear
where $f =1 - 2\mu/R^{\bar d}$,
\beq{2.32}
{\cal F}^s= - \frac{Q_s}{R^{d_1}} (\prod_{s' \in S}  H_{s'}^{- A_{s s'}})
dR \wedge\tau(I_s), \quad s\in S_e,
\eeq
\beq{2.33}
{\cal F}^s= Q_s \tau(\bar I_s), \quad s\in S_m.
\eeq
Here $Q_s \neq 0$ ($s\in S$) are charges, $R_0 >0$,
$R_0^{\bar d} =2 \mu > 0$ and $\bar d = d_1 -1$. In  (\ref{2.30})
$g^i=g_{m_i n_i}^i(y_i) dy_i^{m_i}\otimes
dy_i^{n_i}$ is a Ricci-flat  metric on $M_{i}$, $i=  3,\ldots,n$ and
$\delta_{iI}=  \sum_{j\in I} \delta_{ij}$ is the indicator of $i$ belonging
to $I$: $\delta_{iI}=  1$ for $i\in I$ and $\delta_{iI}=  0$ otherwise.
Here  $g^2 = -dt \otimes dt$, and $g^1 = d \Omega_{d_1}$ is a canonical
metric on unit sphere $M_1 =S^{d_1}$,

The  $p$-brane  set  $S$ is by definition
\beq{1.6}
S = S_e \cup S_m, \quad
S_v = \cup_{a\in\tri}\{a\}\times\{v\}\times\Omega_{a,v},
\eeq
$v = e,m$ and $\Omega_{a,e}, \Omega_{a,m} \subset \Omega$, where $\Omega =
\Omega(n)$  is the set of all non-empty subsets of $\{ 2, \ldots,n \}$,
i.e.
all $p$-branes do not ``live'' in  $M_1$.

Any $p$-brane index $s \in S$ has the form $s = (a_s,v_s, I_s)$, where
$a_s \in \tri$, $v_s =  e,m$ and $I_s \in \Omega_{a_s,v_s}$. The sets $S_e$
and $S_m$ define electric and magnetic $p$-branes, correspondingly. In
(\ref{2.31}) $\chi_s  =   +1, -1$ for $s \in S_e, S_m$, respectively. All
$p$-branes contain the time manifold $M_2 = \R$, i.e.
\beq{1.7a}
2 \in I_s, \qquad \forall s \in S.
\eeq

All manifolds $M_{i}$, $i > 2$, are oriented and connected and
\beq{1.12}
\tau_i  \equiv \sqrt{|g^i(y_i)|} \ dy_i^{1} \wedge \ldots \wedge
dy_i^{d_i},
\eeq
are volume $d_i$-forms, where $d_{i} = {\rm dim} M_{i}$, $i = 1, \ldots,
n$,
with $d_1 > 1$ and $d_2 = 1$. For any $I =   \{ i_1, \ldots, i_k \} \in
\Omega$, $i_1 < \ldots < i_k$, we denote
\beq{1.13}
\tau(I) \equiv \tau_{i_1}  \wedge \ldots \wedge \tau_{i_k}, \qquad
d(I)  =  \sum_{i \in I} d_i.
\eeq
The forms ${\cal F}^s$ correspond to the electric and magnetic $p$-branes
for $s\in S_e, S_m$, respectively. In (\ref{2.33}) we use the notation
$\bar I = \{1,\ldots,n\}\setminus I$.

The parameters  $h_s$ appearing in the solution satisfy the relations:
$h_s = (B_{s s})^{-1}$, where
\beq{1.17}
B_{ss'} = d(I_s\cap I_{s'})+\frac{d(I_s)d(I_{s'})}{2-D}+ \chi_s\chi_{s'}
\lambda_{\alpha a_s}\lambda_{\beta a_{s'}}h^{\alpha\beta},
\eeq
$s, s' \in S$, with $(h^{\alpha\beta})=(h_{\alpha\beta})^{-1}$ and $D = 1 +
\sum_{i =   1}^{n} d_{i}$. Here we assume that

$$({\bf i}) \  B_{ss} \neq 0,$$

$s \in S$, and

$$({\bf ii}) \ {\rm det}(B_{s s'}) \neq 0,$$

i.e. the matrix  $(B_{ss'})$ is a non-degenerate one.

Let us consider the matrix
\beq{1.18}
(A_{ss'}) = \left( 2 B_{s s'}/B_{s' s'} \right).
\eeq
Here  some ordering in $S$ is assumed.

The functions $H_s = H_s(z) > 0$, $z = 2\mu/R^{\bar d} \in (0,1)$,
obey the equations
\beq{3.1}
\frac{d}{dz} \left( \frac{(1-z)}{H_s} \frac{d H_s}{dz} \right) = B_s
\prod_{s' \in S}  H_{s'}^{- A_{s s'}},
\eeq
equipped with the boundary conditions
\bear{3.2a}
H_{s}(1 - 0) = H_{s0} \in (0, + \infty), \\ \label{3.2b}
H_{s}(+ 0) = 1,
\ear
$s \in S$. Here
\beq{1.22}
B_{s} = B_{ss} \eps_s Q_s^2/(2 \bar d \mu)^2, \qquad
\eps_s=(-\eps[g])^{(1-\chi_s)/2}\eps(I_s) \theta_{a_s},
\eeq
$s\in S$, $\eps[g]\equiv\sign\det(g_{MN})$. More explicitly eq.
(\ref{1.22})
reads: $\eps_s=\eps(I_s) \theta_{a_s}$ for $v_s = e$ and $\eps_s=-\eps[g]
\eps(I_s) \theta_{a_s}$, for $v_s = m$.

Eqs. (\ref{3.1}) had to be solved in order to obtain
explicit solutions. They are equivalent to the Toda-type equations
\cite{IMtop}. The first boundary condition (\ref{3.2a})
guarantees the existence of a regular horizon at
$R^{\bar{d}} = 2 \mu$. The second condition (\ref{3.2b}) ensures an
asymptotical flatness (for $R \to + \infty$) of the $(2+d_1)$-dimensional
section of the metric.

Due to eqs. (\ref{2.32}) and (\ref{2.33}), the dimension of $p$-brane
world volume $d(I_s)$ is defined by relations: $d(I_s) =  n_{a_s}-1$,
$d(I_s) = D - n_{a_s} -1$, for $s \in S_e, S_m$, respectively. For a
$p$-brane we use a standard notation: $p = p_s = d(I_s)-1$.

The solutions are valid if the appropriate restriction on the sets
$\Omega_{a,v}$ is imposed. This restriction guarantees the block-diagonal
structure of the stress-energy tensor. We denote $w_1\equiv\{i|i\in
\{2,\dots,n\},\quad d_i=1\}$ and $n_1=|w_1|$ (i.e. $n_1$ is the number of
1-dimensional spaces among $M_i$, $i=2,\dots,n$). It is clear, that
$2 \in w_1$.

{\bf Restriction.} {\em Let 1a) $n_1\le1$ or 1b) $n_1\ge2$ and for any
$a\in\tri$, $v\in\{e,m\}$, $i,j\in w_1$, $i \neq j$, there are no
$I,J\in\Omega_{a,v}$ such that $i \in I$, $j\in J$ and $I\setminus\{i\}=
J\setminus\{j\}$.}

This restriction is  satisfied in the non-composite case: $|\Omega_{a,e}| +
|\Omega_{a,m}| = 1$, (i.e when there are no two  $p$-branes with the same
color index $a$, $a\in\tri$.)  The restriction forbids certain
intersections
of two $p$-branes with the same color index for  $n_1 \geq 2$.

The Hawking temperature corresponding to the solution has the form
\cite{IMp1}
\beq{2.36}
T_H = \frac{\bar{d}}{4 \pi (2 \mu)^{1/\bar{d}}}\prod_{s \in S}
H_{s0}^{- h_s},
\eeq
where $H_{s0}$ are defined in (\ref{3.2a}).

The solution under consideration describes a set of charged (by forms)
overlapping black $p$-branes "living" on submanifolds of $M_2 \times \dots
\times M_n$.

\section{Examples of solutions}
\subsection{``Block-orthogonal'' solutions}

The simplest polynomial solutions of eqs. (\ref{3.1})
occur in orthogonal case \cite{CT,AIV,Oh,BIM,IMJ}, when
\beq{3.4}
B_{s s'} = 0,
\eeq
for  $s \neq s'$, $s, s' \in S$. In this case $(A_{s s'}) = {\rm diag}(2,
\ldots,2)$ is a Cartan matrix for semisimple Lie algebra $A_1 \oplus
\ldots  \oplus  A_1$ and
\beq{3.5}
H_{s}(z) = 1 + P_s z,
\eeq
with $P_s \neq 0$, satisfying
\beq{3.5a}
P_s(P_s + 1) = - B_s,
\eeq
$s \in S$. For positive parameters $P_s > 0$  we get negative $B_s < 0$.

In \cite{Br,IMJ2} this solution  of eqs. (\ref{3.1})
was generalised to the "block-orthogonal" case:
\beq{3.6}
S=S_1 \cup\dots\cup S_k, \qquad  S_i \cap S_j = \emptyset, \quad i \neq j,
\eeq
$S_i \ne \emptyset$, i.e. the set $S$ is a union of $k$ non-intersecting
(non-empty) subsets $S_1,\dots,S_k$, and relation (\ref{3.4}) is satisfied
for all $s\in S_i$, $s'\in S_j$, $i\ne j$; $i,j=1,\dots,k$. In this case
eq. (\ref{3.5}) is modified as follows
\beq{3.8}
H_{s}(z) = (1 + P_s z)^{b^s},
\eeq
where
\beq{3.11}
b^s = 2 \sum_{s' \in S} A^{s s'},
\eeq
$(A^{s s'}) = (A_{s s'})^{-1}$ and parameters $P_s$ are coinciding inside
blocks, i.e. $P_s = P_{s'}$ for $s, s' \in S_i$, $i =1,\dots,k$. Parameters
$P_s \neq 0 $ satisfy the relations (\ref{3.5a}) and parameters $B_s$  are
also  coinciding inside blocks, i.e. $ B_s =  B_{s'}$ for $s, s' \in S_i$,
$i =1,\dots,k$.

Let $(A_{s s'})$ be a Cartan matrix for a finite-dimensional semisimple
Lie algebra $\cal G$. In this case all the powers in (\ref{3.11}) are
natural
numbers coinciding with the components  of twice the dual Weyl vector in
the basis of simple roots \cite{FS}, and  hence, all functions $H_s$ are
polynomials, $s \in S$. The following conjecture may help in finding
solutions   of eqs. (\ref{3.1}).

{\bf Conjecture \cite{IMp1}.} {\em Let $(A_{s s'})$ be  a Cartan matrix
for a  semisimple finite-dimensional Lie algebra $\cal G$. Then  the
solution
to eqs. (\ref{3.1})-(\ref{3.2b}) (if exists) is a polynomial
\beq{3.12}
H_{s}(z) = 1 + \sum_{k = 1}^{n_s} P_s^{(k)} z^k,
\eeq
where $P_s^{(k)}$ are constants, $k = 1,\ldots, n_s$, the integers $n_s =
b^s$ are defined in (\ref{3.11}) and $P_s^{(n_s)} \neq 0$,  $s \in S$}.

This conjecture
was verified for $A_n$ and $C_n$  series of Lie algebras
\cite{IMp2,IMp3}. In the extremal case $\mu = + 0$ an a analogue of this
conjecture was suggested earlier in \cite{LMMP}.

\subsection{Solution for $A_2$ algebra}

Here before coming to solutions for $A_3$ algebra
we present the polynomial solution from \cite{IMp1,IMp2} corresponding
to the Lie algebra $A_2 = sl(3)$ with the Cartan matrix
\beq{B.1a}
\left(A_{ss'}\right)=\left( \begin{array}{*{6}{c}}
2&-1\\
-1&2\\
\end{array}\right).
\eeq

The moduli polynomials  read in this case as follows
\beq{4.1}
H_{s} = 1 + P_s z + P_s^{(2)} z^{2},
\eeq
where $P_s= P_s^{(1)}$ and $P_s^{(2)} \neq 0$ are constants and
\bear{4.5}
P_s^{(2)} = \frac{ P_s P_{s +1} (P_s + 1 )}{2 (P_1 +P_2 + 2)}, \\
\label{4.6}
B_s = - \frac{ P_s (P_s + 1 )(P_s + 2 )}{P_1 +P_2 + 2},
\ear
$s = 1,2$.  Here $P_1 +P_2 + 2 \neq 0$.

In the $A_2$-case the solution is described by relations
(\ref{2.30})-(\ref{2.33}) with $S = \{s_1,s_2\}$ and intersection rules
are following from (\ref{1.17}), (\ref{1.18}) and (\ref{B.1a})
\bear{1.40a}
d(I_{s_1} \cap I_{s_2})= \frac{d(I_{s_1})d(I_{s_2})}{D-2}-
\chi_{s_1} \chi_{s_2} \lambda_{a_{s_1}}\cdot\lambda_{a_{s_2}}
- \frac12 K, \\ \label{1.40b}
d(I_{s_i}) - \frac{(d(I_{s_i}))^2}{D-2}+\lambda_{a_{s_i}}\cdot
\lambda_{a_{s_i}} = K,
\ear
where
$K \neq 0$  and functions $H_{s_i} = H_i$ are defined by relations
(\ref{4.1})-(\ref{4.6}) with $z = 2\mu R^{-\bar d}$, $i =1,2$. Here and in
what follows $\lambda\cdot\lambda^{'}=\lambda_{\alpha}\lambda_{\beta }^{'}
h^{\alpha\beta}$.

\subsection{Solution for $A_3$ algebra}

Now we present the new
solution related to the Lie algebra $A_3 =sl(4)$ with
the Cartan matrix
\beq{B.1c}
\left(A_{ss'}\right)=\left(\begin{array}{ccc}
2 & -1 & 0 \\
-1 & 2 & -1 \\
0 & -1 & 2
\end{array}\right).
\eeq

According to {\bf Conjecture} we  seek the solution to eqs.
(\ref{3.1})-(\ref{3.2b}) in the following form
\bear{5.1}
H_1(z)=1+P_1z+P_1^{(2)}z^2+P_1^{(3)}z^3, \\ \label{5.2}
H_2(z)=1+P_2z+P_2^{(2)}z^2+P_2^{(3)}z^3+P_2^{(4)}z^4, \\
\label{5.3a}
H_3(z)=1+P_3z+P_3^{(2)}z^2+P_3^{(3)}z^3,
\ear
where $P_s= P_s^{(1)}$ and $P_s^{(k)}$ are constants, $s = 1,2,3$.

Here we outline the result obtained by computer calculations.
For $B_s$-parameters (\ref{1.22}) we get the following relations
\bear{5.3}
B_1=\frac1{2(2 + P_1 + P_2 - P_3)}\Bigl[12 - \Delta - 2P_1^3 -
2P_1^2(3 - P_3) + 6P_3 \\ \nn
+ P_2(4 + P_2)(3 + P_3) + P_1(2 + P_2(4 + P_2) + 6P_3)\Bigr],
\\ \label{5.4}
B_3=\frac1{2(2 - P_1 + P_2 + P_3)}\Bigl[12 - \Delta + 6P_1 +
(3 + P_1)P_2(4 + P_2), \\ \nn
+ (2 + 6P_1 + P_2(4 + P_2))P_3 - 2(3 - P_1)P_3^2 - 2P_3^3\Bigr],
\\ \label{5.5a}
B_2=\frac1{(2 + P_1 + P_2 - P_3)(2 - P_1 + P_2 + P_3)}\Bigl[(2 + P_2)
(\Delta \\ \nn
- (2 + P_2)(2 + (2 + P_2)^2 + 2P_1P_3 + 3(P_1 + P_3)))\Bigr],
\ear
and for parameters $P_s^{(k)}$ in (\ref{3.1})-(\ref{3.2b})
we obtain
\bear{5.5}
P_1^{(2)}=\frac1{4(2 + P_1 + P_2 - P_3)}\Bigl[12 - \Delta + 2P_1^2P_2 +
6P_3 + P_2(4 + P_2)(3 + P_3) \\ \nn
+ P_1(6 + P_2(6 + P_2) + 4P_3)\Bigr], \\  \label{5.6}
P_3^{(2)}=\frac1{4(2 - P_1 + P_2 + P_3)}\Bigl[12 - \Delta + 6P_1 + (3 +
P_1)P_2(4 + P_2) \\ \nn
+ (6 + 4P_1 + P_2(6 + P_2))P_3 + 2P_2P_3^2\Bigr], \\  \label{5.7}
P_1^{(3)}=\frac1{12(2 + P_1 + P_2 - P_3)}\Bigl[(2 + P_1 + P_2)(12 -
\Delta + 6P_1 + (3 + P_1)P_2(4 + P_2)) \\ \nn
+ (2P_1^2(2 + P_2) + (2 + P_2)(6 + P_2(4 + P_2)) + P_1(14 + P_2(10 +
P_2)))P_3\Bigr], \\  \label{5.8}
P_3^{(3)}=\frac1{12(2 - P_1 + P_2 + P_3)}\Bigl[P_1((2 + P_2)(6 + P_2(4 +
P_2)) + (14 + P_2(10 + P_2))P_3 \\ \nn
+ 2(2 + P_2)P_3^2) + (2 + P_2 + P_3)(12 - \Delta + 6P_3 + P_2(4 + P_2)
(3 + P_3))\Bigr], \\  \label{5.9}
P_2^{(2)}=\frac1{2(2 + P_1 + P_2 - P_3)(2 - P_1 + P_2 + P_3)}\Bigl[-((2 +
P_2)(12 - \Delta + 3P_2(4 + P_2)) \\ \nn
+ 2P_1(4 + 3P_2) P_3 + 3(2 + P_2)^2(P_1 + P_3) + P_2(1 + P_2)(P_1^2 +
P_3^2))\Bigr],
\\
P_2^{(3)}=\frac1{6(2 + P_1 + P_2 - P_3)(2 - P_1 + P_2 + P_3)}\Bigl[(2 +
P_2)(-(P_2^3(3 + P_1 + P_3)) \\ \nn
+ \Delta(2 + P_1 + P_2 + P_3) - P_2^2(18 + P_1(9 + P_1) + P_3(9 + P_3)) -
2(2 + P_1 + P_3) \\ \nn
\times(6 + 2P_1P_3 + 3(P_1 + P_3) + P_2(9 + P_1P_3 + 2(P_1 + P_3))))\Bigr],
\\
P_2^{(4)}=\frac1{24(2 + P_1 + P_2 - P_3)(2 - P_1 + P_2 + P_3)}\Bigl[(2 +
P_2)(-((3 + P_2)(2 + P_2 + P_3) \\ \nn
\times(12 - \Delta + 6P_3 + P_2(4 + P_2)(3 + P_3))) + P_1((-2 - P_2)(3 +
P_2)^2(4 + P_2) \\ \nn
- 2(42 + P_2(42 + P_2(12 + P_2)))P_3 - 2(2 + P_2)(6 + P_2)P_3^2 +
\Delta(3 + P_2 + 2P_3)) \\ \nn
- P_1^2(P_2^3 + 2(3 + 2P_3)^2 + P_2^2(7 + 2P_3) + P_2(18 + 4P_3(4 +
P_3))))\Bigr],
\ear
where
\bear{5.10}
\Delta=\biggl[P_1^2\left((6 + P_2(4 + P_2))^2 + 12(2 + P_2)^2P_3 +
4(2 + P_2)^2P_3^2\right) \\ \nn
+ 2P_1\left(3(2 + P_2)^2\left(6 + P_2(4 + P_2)\right) + \left(84 + P_2(4 +
P_2)\left(26 + P_2(4 + P_2)\right)\right)P_3 \right. \\ \nn \left.
+ 6(2 + P_2)^2P_3^2\right) + \left(6(2 + P_3) + P_2(4 + P_2)(3 +
P_3)\right)^2\biggr]^{1/2}.
\ear
Here $2 +  P_2  \pm (P_1 - P_3) \neq 0$.

The $A_3$ black-brane solution is described by relations
(\ref{2.30})-(\ref{2.33}), with $S = \{s_1,s_2,s_3\}$, and intersection
rules following from (\ref{1.17}), (\ref{1.18}) and (\ref{B.1c})
\bear{5.11a}
d(I_{s_1} \cap I_{s_2})=\frac{d(I_{s_1})d(I_{s_2})}{D-2}- \chi_{s_1}
\chi_{s_2}\lambda_{a_{s_1}}\cdot\lambda_{a_{s_2}} - \frac12 K,
\\ \label{5.11b}
d(I_{s_2} \cap I_{s_3})=\frac{d(I_{s_2})d(I_{s_3})}{D-2}- \chi_{s_2}
\chi_{s_3}\lambda_{a_{s_2}}\cdot\lambda_{a_{s_3}} - \frac12 K,
\\ \label{5.11c}
d(I_{s_1} \cap I_{s_3})=\frac{d(I_{s_1})d(I_{s_3})}{D-2}- \chi_{s_1}
\chi_{s_3}\lambda_{a_{s_1}}\cdot\lambda_{a_{s_3}},
\\ \label{5.11d}
d(I_{s_i}) - \frac{(d(I_{s_i}))^2}{D-2}+\lambda_{a_{s_i}}\cdot
\lambda_{a_{s_i}} = K,
\ear
where $K \neq 0$  and functions $H_{s_i} = H_i$ are defined by relations
(\ref{5.1})-(\ref{5.3a}) with $z = 2\mu R^{-\bar d}$, $i =1,2,3$.

For $P_1 = 3P$, $P_2 = 4P$ and $P_3 = 3P$ with $P > 0$
we get a special ``block-orthogonal'' solution
\beq{5.1b}
H_1(z)= (1+ P z)^3, \quad  H_2(z)= (1+ P z)^4, \quad
H_3(z)= (1+ P z)^3,
\eeq
in agreement with the relations (\ref{3.8}) and (\ref{3.11}).

In the special case $P_1= P_3$ the relations for $H_1$, $H_2$
(\ref{5.1})-(\ref{5.2})   describe the solution corresponding to Lie
algebra $C_2 = so(5)$ \cite{GIK}.

\subsubsection{Solution with three $3$-branes in $D=12$ model}

Now we illustrate the general $A_3$ formulae by considering a bosonic field
model in dimension  $D= 12$ \cite{KKLP} that admits the bosonic sector
of 11-dimensional supergravity as a consistent truncation. The action
for this model has the following form
\bear{a.1}
S_{12} =
\int_{M} d^{12}z \sqrt{|g|} \{ {R}[g] -
g^{MN} \partial_{M} \varphi \partial_{N} \varphi
- \frac{1}{4!} \exp( 2 \lambda_{4} \varphi)  (F^4)^2
- \frac{1}{5!} \exp( 2 \lambda_{5} \varphi ) (F^5)^2 \}
\\ \nn
+ c_{12} \int_{M} A^4 \wedge F^4 \wedge F^4.
\ear
Here  $F^4 = dA_3$ is the $4$-form,
$F^5 = dA^4$ is the $5$-form, $c_{12}$ is constant and
\beq{a.2}
 \lambda_{4}^2 = - \frac{1}{10}, \quad
 \lambda_{5} = - 2 \lambda_{4}.
\eeq

The model describes electrically charged 2- and 3-branes and
magnetically charged 6- and 5-branes (corresponding to $F^4$ and $F^5$,
respectively) with (orthogonal) intersection rules given in \cite{IMC}.

First, we consider the $A_3$-solution for the truncated model
(i.e. without Chern-Simons term) defined on the manifold
\bear{6.2}
M = (2 \mu, + \infty) \times (M_1 = S^{2}) \times (M_2 = \R)
\times (M_3 = \R^2) \\ \nn
\times (M_4 = \R^3) \times (M_5 = \R^2) \times (M_6 = \R),
\ear
with three electric 3-branes with sets $I_i = I_{s_i}$ defined
as follows
\beq{6.3}
I_1 = \{ 2, 3, 6\}, \qquad  I_2 = \{ 2, 4 \}, \qquad
I_3 = \{ 2, 5, 6\},
\eeq
The $A_3$-intersection rules (\ref{5.11a})-(\ref{5.11d})
read
\beq{6.4}
d(I_1 \cap I_2) = d(I_2 \cap I_3) = 1, \qquad
d(I_1 \cap I_3) = 2,
\eeq
(here $K = 2$). The black-brane solution has the following form
\bear{6.30}
g= (H_1 H_2 H_3)^{4/10} \biggl\{ f^{-1} dR \otimes dR +
                                R^2  d \Omega^2_{2}  \\ \nn
- (H_1 H_2 H_3)^{-1} f dt \otimes dt
+  H_1^{-1} g^3  + H_2^{-1} g^4  + H_3^{-1} g^5
+ H_1^{-1} H_3^{-1} g^6  \biggr\},
\\ \label{6.31}
\exp(\varphi)= (H_1 H_2 H_3)^{\lambda_5/2},
\\  \label{6.32a}
F^5=
\frac{Q_1}{R^2}H_1^{- 2} H_2 dt \wedge dR \wedge \tau_3 \wedge \tau_6
+ \frac{Q_2}{R^2}H_1H_2^{- 2} H_3 dt \wedge dR \wedge \tau_4 +
\\ \nn
\frac{Q_3}{R^2}H_2 H_3^{- 2} dt \wedge dR \wedge \tau_5 \wedge \tau_6,
\ear
and $F^4 = 0$, where $f =1 - 2\mu/R$.

Here $Q_i \neq 0$ are charges related to parameters $P_i$,
by formulae $B_i = - Q_i^2/(2 \mu^2)$,
 and (\ref{5.3})-(\ref{5.5a}); $\mu > 0$,
and the functions $H_i = H_i(z, (P_j))$ are defined by
(\ref{5.1})-(\ref{5.3a}), (\ref{5.5})-(\ref{5.10}),
and
$z = 2\mu/R$, $i =1,2,3$. In (\ref{6.30}) $g^a$
is a flat metric on $M_a$, $a = 3,\ldots,6$.

This solution satisfies the equations of motion for the  model
(\ref{a.1}) itself,
since the only modifications due to Chern-Simons term are related to
``Maxwells'' equations
\ber{4.14}
d*F^4 = {\rm const} \ F^5 \wedge F^4, \qquad
d*F^5 = {\rm const} \ F^4 \wedge F^4,
\eer
and are trivial (since $F^4 = 0)$.

We remind that the twelve-dimensional
model from \cite{KKLP}
can  be consistently truncated to a
ten-dimensional Lagrangian that contains all the $BPS$ $p$-brane solutions
of the type $IIB$-theory.

\section{Conclusions}

In this paper we presented a new black-brane solution with three branes
and intersection rules corresponding to the Lie algebra $A_3$. The solution
is governed by polynomials of degree $3, 4, 3$ and the coefficients of
polynomials exhibit a special dependence upon the parameters $P_1$,
$P_2$ and $P_3$ due to appearance of the term $\Delta$ (see
(\ref{5.10})). (The case $P_1= P_3$, corresponding to Lie algebra $C_2
= so(5)$ was considered recently in \cite{GIK}.)

The $A_3$-solution is more complicated then the $A_2$-one, which has a
rather simple analytical structure. This means that the polynomial
solutions for other $A_n$-algebras ($n > 3$) may be
even more complicated.

Other topics of interest is the application of this result
to black-brane thermodynamics
(e.g. relations for the entropy,  Cardy-Verlinde
type formulas etc) and analysis of post-Newtonian effects \cite{IMp2}.
On this way one may expect to clarify the appearance of the function
$\Delta$ in the solution. Another problem is to find the polynomial
solutions for other Lie algebras (at least for $A_k$ ones).

Here we also
considered as an example the $A_3$-solution with three $3$-branes in
$12$-dimensional model from \cite{KKLP} (this model
imitates the low-energy limit of hypothetical $F$ theory \cite{Vafa}).
This solution may be also generalized to the case of $B_D$-models
\cite{IMJ} in dimensions $D \geq 12$.

\begin{center}
{\bf Acknowledgments}
\end{center}

This work was supported in part by the Russian Ministry
of Industry, Science and Technology, Russian Foundation for Basic Research
(RFFI-01-02-17312-a) and DFG project (436 RUS 113/678/0-1(R)).
The results of the work were reported at the
Fifth International Conference on Gravitation
and Astrophysics of Asian-Pacific countries.

\end{document}